\documentclass[pdflatex]{revtex4}

\usepackage{latexsym}
\usepackage{amsmath,amssymb,bm,empheq} 
\usepackage[pdftex]{graphicx} 
\usepackage[pdftex, colorlinks=true, linkcolor=red,citecolor=blue]{hyperref} 

\makeatletter

\newcommand{\cket}[1]{| #1 \rangle} 
\newcommand{\bra}[1]{\langle #1 |} 
\newcommand{\ImUnit}{\mathrm{i}} 
\newcommand{\evec}[1]{\boldsymbol{#1}} 
\newcommand{\tr}{\mathrm{Tr}} 

\begin{document}

\title{Is quantum teleportation beyond horizon possible?}

\author{Jun-ichirou Koga} 
\affiliation{Faculty of Science and Engineering, 
Waseda University, Tokyo 169-8555, Japan} 
\email{koga@waseda.jp} 

\author{Kengo Maeda} 
\affiliation{Faculty of Engineering, 
Shibaura Institute of Technology, Saitama 330-8570, Japan} 
\email{maeda302@sic.shibaura-it.ac.jp}
 
\begin{abstract} 
We ask whether quantum teleportation from the outside to the inside of a horizon is possible, using entanglement extracted from a vacuum. 
We first calculate analytically, within the perturbation theory, entanglement extracted from the Minkowski vacuum into a pair of an inertial and an accelerated Unruh-DeWitt detectors, which are initially in the ground states and interact with a neutral massless scalar field for an infinitely long time. 
We find that entanglement can be extracted, but is ``fragile'', depending on adiabaticity of switching of the detectors at the infinite past and future. 
We then consider the standard scheme of quantum teleportation utilizing the extracted entanglement, and find that the standard teleportation is not superior to channels without entanglement. 
\end{abstract}

\maketitle 

\section{Introduction} 

The information problem of black hole \cite{Hawking76} still remains as a great challenge in theoretical physics, in spite of vigorous interdisciplinary researches. See, e.g., Refs.\cite{UnruhWald17,AlmheiriHMST20} for recent review articles. In particular, quantum information theory plays crucial roles in recent developments, and quantum teleportation \cite{BennettBCJPW93} is one of the cornerstones in quantum information, with which a sender, usually named Alice, can transmit an unknown quantum state to a remote receiver, Bob, based on quantum entanglement between them. 

Since quantum teleportation involves classical communication as a part of its task, which is of course subject to causality, quantum teleportation from the inside to the outside of a black hole is impossible, in principle. Conversely, can one teleport a quantum state from the outside to the inside of a black hole? 

It seems that many attempts so far to resolve the information problem are based on strong entanglement between the inside and the outside of a black hole. If they are  strongly entangled, which is used to retrieve the information gravitationally fallen into the black hole and emitted outside as Hawking radiation \cite{Hawking75}, it is conceivable  that such strong entanglement might enable one to send the emitted information back into the black hole through quantum teleportation. Or, one might send by quantum teleportation information that has remained outside during the evolution into a black hole more efficiently than free fall. It is then interesting to see whether these processes are indeed possible. 

As the first step to analyze this issue, however, entanglement between the inside and the outside of a black hole is not necessarily prepared initially, since one can extract entanglement from a vacuum \cite{SummersWerner} into a pair of qubits interacting with a quantum field \cite{Reznik03}. If one can implement quantum teleportation without pre-existing entanglement, it will be possible also when entanglement is prepared initially. 
Thus, it will be meaningful enough to ask whether or not quantum teleportation using only entanglement extracted from a vacuum is efficiently performed from the outside to the inside of a black hole. 
To see this, we will consider in this paper an inertial and an accelerated observer in the Minkowski spacetime, which correspond to an observer free falling into a black hole and a static observer outside a black hole, respectively, due to the equivalence principle. Each of these observers is assumed to hold a qubit, and we will consider quantum teleportation sufficiently later than the instant when the inertial observer goes across the Rindler horizon of the accelerated observer.  

As a model of qubits that interact with a quantum field, we will consider two-level Unruh-DeWitt detectors \cite{Unruh76,DeWitt79,BirrellDavies82} interacting with a neutral massless scalar field in the Minkowski vacuum. On the one hand, it is well-known that a single Unruh-DeWitt detector provides an operationally well-defined definition of quantum particles. In particular, in the Minkowski vacuum, an inertial Unruh-DeWitt detector does not detect any particles while an accelerated Unruh-DeWitt detector responds as if it is immersed in a thermal bath, the so-called Unruh effect \cite{Unruh76}. We emphasize here that these properties of the Unruh-DeWitt detector result from infinitely long interaction with a quantum field to be probed. On the other hand, a pair of two-level Unruh-DeWitt detectors has been used as a tool to extract entanglement from a vacuum, which is often called entanglement ``harvesting'' \cite{SaltonMM15} and has been applied even to black hole spacetimes \cite{HerdersonHMSZ18,TjoaMann20,GallockYoshimuraTM21-}, where the interaction between the detectors and the quantum field is usually assumed to vary in time according to the Gaussian profile. 

So far, quantum teleportation from an {\it inertial} observer to an {\it accelerated} observer in the Minkowski spacetime has been considered \cite{AlsingMilburn03,LandulfoMatsas09,LinCH15}, where the Rindler horizon is absent due to a finite period of acceleration or the inertial observer sends a classical message before crossing the Rindler horizon. On the other hand, quantum teleportation in the opposite direction, i.e., from an accelerated observer to an inertial observer,  was considered very recently in Ref. \cite{FooRalph20} in the context of continuous variable scheme, rather than qubits, while involving a beam-splitter with time-dependent reflectivity in the future Rindler wedge. It is shown in Ref. \cite{FooRalph20} that the output state obtained by the inertial observer is polluted by thermal noise. The introduction of the time-dependent reflectivity of the beam-splitter was motivated from the quantization in the future Rindler wedge \cite{OlsonRalph11,HiguchiIUK17}, which is performed with respect to a timelike {\it conformal} Killing vector, rather than a timelike Killing vector, and thus a detector that probes the quantum particles in the future Rindler wedge will need to have time-dependent excitation energy \cite{OlsonRalph11}, or equivalently it should expand as time elapses\cite{Wald19}. 

We will then consider in this paper rigid (non-expanding) point-like Unruh-DeWitt detectors that follow the worldlines, to which timelike Killing vectors are tangent. They are assumed to interact with the quantum scalar field for an infinitely long time and are switched on and off sufficiently adiabatically at the infinite past and future, respectively. We mention here that not only a single Unruh-DeWitt detector probes the quantum state faithfully, but also that a pair of Unruh-DeWitt detectors interacting with a quantum field for an infinitely long time is a useful tool to investigate entanglement of the state of a quantum field. On the one hand, they are not entangled when they are inertial and comoving \cite{KogaKM18}, which thus provides a fiducial system to quantify entanglement. On the other hand, they are properly entangled when they are uniformly accelerated in the opposite directions in the left and right Rindler wedges, respectively, which are known to be strongly entangled in the Minkowski vacuum \cite{Reznik03,KogaMK19}.

\section{Reduced density matrix of detectors} 

We consider a pair of two-level Unruh-DeWitt detectors in the Minkowski spacetime, $A$ possessed by Alice and $B$ by Bob, which are prepared in the ground states at infinite past $t \rightarrow - \infty$. These two detectors, denoted collectively as $I$, where $I = A , B$, are assumed to interact with a neutral massless scalar field $\phi(x)$ through the interaction Lagrangian $L_{\rm int}$, whose time integral, the interaction action $\mathcal{S}_{{\rm int}}$, is given as 
\begin{equation}
\int L_{\rm int} \, d t = \mathcal{S}_{{\rm int}} 
= \int c \, \chi_A(\tau_A) \, m_A(\tau_A) \, \phi(\bar{x}_A) d \tau_A 
+ \int c \, \chi_B(\tau_B) \, m_B(\tau_B) \, \phi(\bar{x}_B) d \tau_B , 
\label{eqn:IntAction} 
\end{equation}
where $t$ is the inertial time coordinate and $c$ is the coupling constant. The proper time of the detector $I$ is denoted as $\tau_I$, 
$\bar{x}_I^{\mu}(\tau_I)$ is the worldline coordinates, and 
$m_I(\tau_I)$ is the monopole operator of the detector $I$. The monopole moment $m_I(\tau_I)$ of each detector is assumed to commute with that of the other detector and with the scalar field $\phi(x)$. The switching function $\chi_I(\tau_I)$ describes how the switching of the detectors is performed. 
In this paper, we focus on the case where the switching is performed adiabatically enough at the infinite past and future, and the Unruh-DeWitt detectors interact with the quantum scalar field $\phi(x)$ for an infinitely long time, as in the textbooks \cite{DeWitt79,BirrellDavies82} on Unruh-DeWitt detectors. 

We further assume that the initial state of the scalar field $\phi(x)$ is the Minkowski vacuum $\cket{0}$ at the infinite past $t \rightarrow - \infty$. Thus, the initial state of the whole system, composed of the two Unruh-DeWitt detectors and the scalar filed, is prepared as $\cket{E_0^{(A)}} \cket{E_0^{(B)}} \cket{0}$, where $\cket{E_0^{(I)}}$ denotes the ground state of the two-level detector $I$, and similarly the excited state of the detector $I$ is denoted as $\cket{E_1^{(I)}}$.  

The standard perturbation theory up to order of $c^2$ gives the reduced density matrix $\rho_{AB}$ of the two Unruh-DeWitt detectors in the infinite future $t \rightarrow \infty$, after tracing over the states of the scalar field, as 
\begin{equation} 
\rho_{AB} = \begin{pmatrix} 
0 & 0 & 0 & c^2 \: \mathcal{E} \\ 
0 & c^2 \: \mathcal{P}_A & c^2 \: \mathcal{P}_{AB} & c^2 \: \mathcal{W}_A \\ 
0 & c^2 \: \mathcal{P}_{AB}^* & c^2 \: \mathcal{P}_B & c^2 \: \mathcal{W}_B \\ 
c^2 \: \mathcal{E}^* & c^2 \: \mathcal{W}_A^* & c^2 \: \mathcal{W}_B^* & 
1 - c^2 \big( \mathcal{P}_A + \mathcal{P}_B \big) \end{pmatrix} + \mathcal{O}(c^4) , 
\label{eqn:DensityMatrixAB} 
\end{equation} 
in a matrix representation with the order of the bases as $\left\{\cket{E_1^{(A)}} \cket{E_1^{(B)}}  , \, 
\cket{E_1^{(A)}} \cket{E_0^{(B)}} , \, \cket{E_0^{(A)}} \cket{E_1^{(B)}} , \, \cket{E_0^{(A)}} \cket{E_0^{(B)}} \right\}$  \cite{KogaKM18} . 
The explicit forms of the components of $\rho_{AB}$ relevant below are given as  
\begin{equation} 
\mathcal{P}_I 
=  \left| \bra{E_{1}^{(I)}} \, m_I(0) \, \cket{E_0^{(I)}} \right|^2 \, \mathcal{I}_I , \quad 
\mathcal{E} 
= \bra{E_{1}^{(B)}} m_B(0) \cket{E_0^{(B)}} \, \bra{E_{1}^{(A)}} m_A(0) \cket{E_0^{(A)}} \, 
\mathcal{I}_E , 
\label{eqn:IntFactorsPE} 
\end{equation} 
where $\mathcal{I}_I$ and $\mathcal{I}_E$ are defined as 
\begin{align}  & 
\mathcal{I}_I \equiv 
\int_{- \infty}^{\infty} d \tau'_I \; \int_{- \infty}^{\infty} d \tau_I \; \chi_I(\tau'_I) \: \chi_I(\tau_I) \: 
e^{\ImUnit \, \Delta E^{(I)} \left( \tau_I -  \tau'_I \right)} \: G_W(\bar{x}'_I , \bar{x}_I) , 
\label{eqn:CalIIDef} \\ & 
\mathcal{I}_E \equiv
- \:  \ImUnit \, \int_{- \infty}^{\infty} d \tau_B \, \int_{- \infty}^{\infty} d \tau_A \:  \chi_B(\tau_B) \, \chi_A(\tau_A) \, 
e^{\ImUnit \, \Delta E^{(B)} \tau_B} e^{\ImUnit \, \Delta E^{(A)} \tau_A} \: G_F(\bar{x}_B , \bar{x}_A) , 
\label{eqn:CalIEDef}
\end{align} 
and $G_W(x,x')$ and $G_F(x,x')$ are the Wightman function and the Feynman propagator of the neutral massless scalar filed in the Minkowski spacetime, respectively, derived as  
\begin{align} & 
G_W(x, x') = \frac{- 1}{( 2 \pi )^2} 
\frac{1}{( t - t' - \ImUnit \, \varepsilon )^2 - | \evec{x} - \evec{x}' |^2} , 
\label{eqn:WightmanMinScalar} \\ & 
G_F(x , x') = \frac{\ImUnit}{( 2 \pi )^2} 
\frac{1}{(t - t')^2 - | \evec{x}' - \evec{x} |^2 - \ImUnit \varepsilon} . 
\label{eqn:FeymnamProExplicit} 
\end{align} 

\begin{figure} 
\includegraphics[scale=0.5,keepaspectratio]{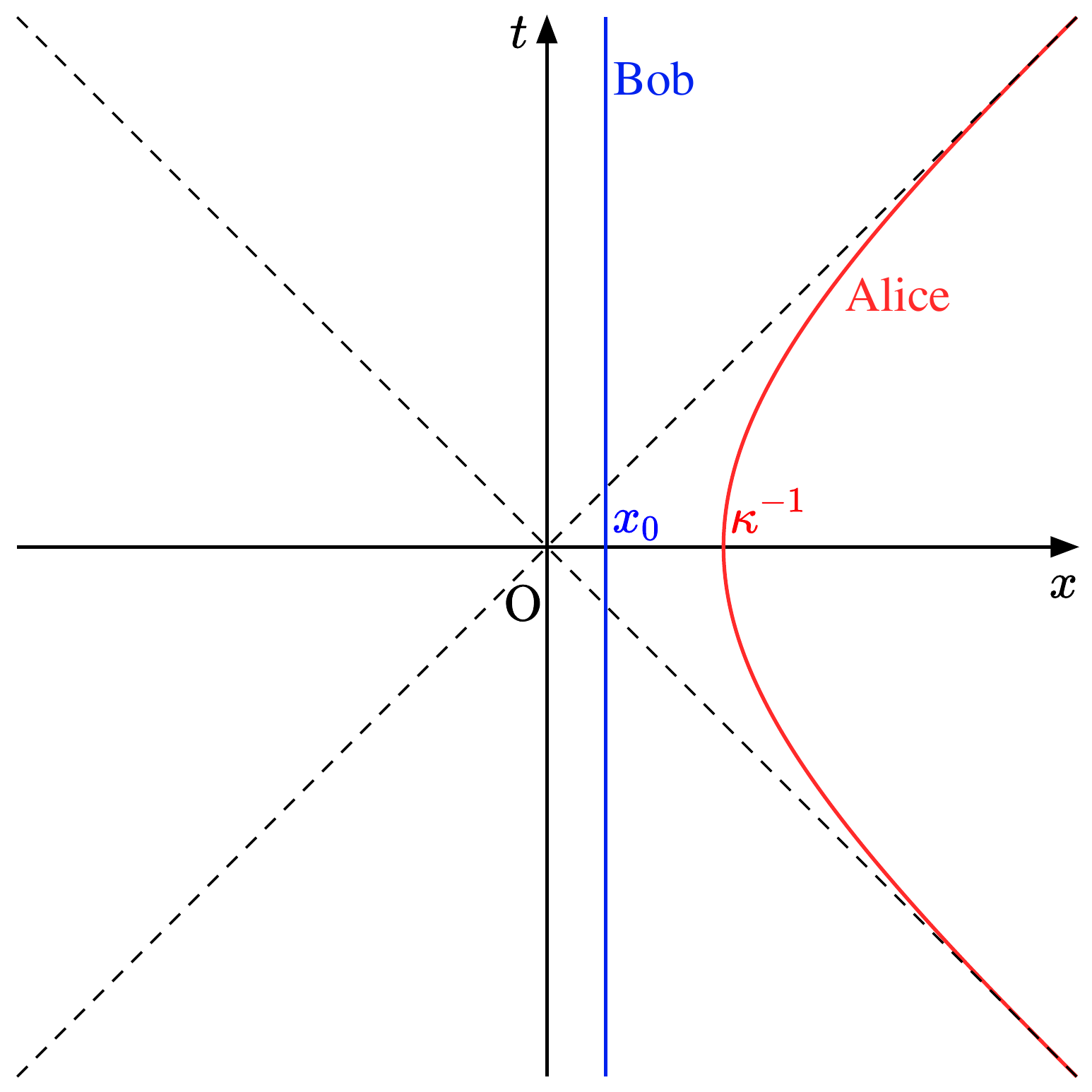} 
\caption{The worldlines of Alice and Bob. At the closest approach, which occurs at $\tau_A = 0$, the $x$-coordinate of Alice is $\bar{x}_A(0) = \kappa^{-1}$. The dashed lines denote the future and past Rindler horizons of Alice.} 
\label{fig:Worldlines}
\end{figure}

Now we specify the worldlines of Alice and Bob. Here we consider the case where Alice is uniformly accelerated 
in the direction of $x$, 
with the proper acceleration $\kappa$, and Bob is at rest at 
$\boldsymbol{x}_0 = ( x_0 , 0 , 0 )$ in the Cartesian coordinate system. 
See FIG. \ref{fig:Worldlines}. 
Their worldline coordinates $\bar{x}_I(\tau_I)$ are thus written as  
\begin{align} & 
\bar{t}_A(\tau_A) = \frac{1}{\kappa} \sinh \left( \kappa \tau_A \right) , \qquad 
\bar{x}_A(\tau_A) = \frac{1}{\kappa} \cosh \left( \kappa \tau_A \right) , \qquad \bar{y}_A(\tau_A) = 0 , \qquad \bar{z}_A(\tau_A) = 0 , 
\notag \\ & 
\bar{t}_B(\tau_B) = \tau_B , \qquad 
\bar{x}_B(\tau_B) = x_0 , \qquad 
\bar{y}_B(\tau_B) = 0 , \qquad \bar{z}_B(\tau_B) = 0 . 
\label{eqn:WorldlinesAliceAccelBobInert} 
\end{align} 
We here assume $0 < x_0 < \kappa^{-1}$, bearing in mind an observer freely falling into a black hole. This restriction also avoids collision of Alice and Bob, which may give rise to an infinite amount of entanglement and invalidate the perturbative analysis. 

By substituting Alice's worldline in Eq. (\ref{eqn:WorldlinesAliceAccelBobInert}) into Eq. (\ref{eqn:CalIIDef}), $\mathcal{I}_A$ for an infinite interaction time $T \rightarrow \infty$ is found to be given by, 
\begin{equation} 
\mathcal{I}_A = \lim_{T \rightarrow \infty} \frac{\Delta E^{(A)}}{2 \pi} \frac{1}{e^{2 \pi \frac{\Delta E^{(A)}}{\kappa}} - 1} \int^T_{- T} d \tau_A 
= \frac{1}{\pi} \, \lim_{T \rightarrow \infty} \, \frac{\Delta E^{(A)}}{e^{2 \pi \frac{\Delta E^{(A)}}{\kappa}} - 1} \: T , 
\label{eqn:CalPAAABI} 
\end{equation} 
where $\Delta E^{(I)}$ is the excitation energy from the ground state to the excited state of the detector $I$, which is assumed to be positive definite $\Delta E^{(I)} > 0$. 
Eq. (\ref{eqn:CalPAAABI}) gives the well-known Planckian distribution of the excitation probability $c^2 \, \mathcal{P}_A$, or the excitation rate $c^2 \dot{\mathcal{P}}_A$ when differentiated with respect to the proper time $\tau_A$, by using Eq. (\ref{eqn:IntFactorsPE}). 
While we set as $\chi_A(\tau_A) = 1$ to derive this expression, we implicitly assume that the switching is performed adiabatically enough, as is usually done. 
Thus, Eq. (\ref{eqn:CalPAAABI}) is expected to approximate well the excitation probability even when we take into account explicitly the effects of a finite (but sufficiently long) interaction time and adiabatic switching. 
Although $\mathcal{I}_A$ in Eq. (\ref{eqn:CalPAAABI}) linearly diverges in $T$ formally, it is meaningful within the validity of the perturbation theory $c^2 \, T \ll 1$, much like Fermi's golden rule in quantum mechanics, 
and we recall also that it collaborates the well-known Unruh effect from the operational standpoint. 

Similarly, the cross-correlation $\mathcal{E}$ will be approximated well by computation with $\chi_I(\tau_I) = 1$, even when we consider the effects of a finite interaction time and adiabatic switching. We thus substitute Eq. (\ref{eqn:WorldlinesAliceAccelBobInert}) into Eq. (\ref{eqn:CalIEDef}) with $\chi_I(\tau_I) = 1$. Then, by following Ref. \cite{KogaMK19}, the integration by $\tau_A$ is performed by considering the infinite semicircle in the upper half of the complex $\tau_A$ plane, which gives 
\begin{equation} 
\mathcal{I}_E 
= \frac{\ImUnit}{4 \pi K} \frac{1}{\kappa} \:  \int_{- \infty}^{\infty} d \tau_B \, 
\frac{e^{\ImUnit \, \Delta E^{(B)} \tau_B}}{D(\tau_B)} 
\left\{ \sum_n e^{\ImUnit \Delta E^{(A)} \tau_{A+n}} - \sum_n e^{\ImUnit \Delta E^{(A)} \tau_{A-n}} \right\} , 
\label{eqn:CalIEFirstIntGen} 
\end{equation} 
where $\tau_{A\pm n}$ is the poles of the Feynman propagator $G_F(\bar{x}_B,\bar{x}_A)$, derived as 
\begin{align} & 
\tau_{A + n} = \left\{ \begin{array}{lll} 
- \kappa^{-1} \ln \kappa \left( x_0 - \tau_B \right) + 2 n \pi \kappa^{-1} \, \ImUnit + \ImUnit \varepsilon 
& \quad & \tau_B < x_0 \\ 
- \kappa^{-1} \ln \kappa \left( \tau_B - x_0 \right) + ( 2 n + 1 ) \pi \kappa^{-1} \, \ImUnit + \ImUnit \varepsilon 
& \quad & x_0 < \tau_B 
\end{array} \right. 
\label{eqn:TauAPNAAccBint1} \\ & 
\tau_{A - n} = \left\{ \begin{array}{lll} 
\kappa^{-1} \ln \kappa \left( - \tau_B - x_0 \right) + ( 2 n + 1 ) \pi \kappa^{-1} \, \ImUnit - \ImUnit \varepsilon 
& \quad & \tau_B < - x_0 \\ 
\kappa^{-1} \ln \kappa \left( x_0 + \tau_B \right) + 2 n \pi \kappa^{-1} \, \ImUnit - \ImUnit \varepsilon 
& \quad & - x_0 < \tau_B 
\end{array} \right. , 
\label{eqn:TauAMNAAccBint1}
\end{align} 
and $\kappa$ and $D(\tau_B)$ are derived in the present case, as 
\begin{equation}
K = \frac{1}{\kappa^2} , \quad D(\tau_B) = \frac{1}{2} \left( 1 - \kappa^2 x_0^2 + \kappa^2 \tau_B^2 \right) . 
\end{equation}
Then, $\mathcal{I}_E$ is written as  
\begin{align} & 
\mathcal{I}_E 
= \frac{1}{8 \pi \sqrt{1 - \kappa^2 x_0^2} \, \sinh \left( \pi \varpi_A \right)} 
\Bigg[ 
e^{\ImUnit \, \Delta E^{(B)} x_0} \, e^{\pi \varpi_A} 
\Big\{ I(\xi_{1+},- \varpi_B,-\varpi_A) 
- I(\xi_{1-},-\varpi_B,-\varpi_A)
\Big\}
\notag \\ & 
- e^{- \ImUnit \, \Delta E^{(B)} x_0} \, e^{- \pi \varpi_A} \, 
\Big\{ I(\xi_{1+},\varpi_B,\varpi_A) 
- I(\xi_{1-},\varpi_B,\varpi_A) 
\Big\}
\notag \\ & 
+ e^{\ImUnit \, \Delta E^{(B)} x_0} \,  
\Big\{ I(\xi_{2+},\varpi_B,- \varpi_A) 
- I(\xi_{2-},\varpi_B,-\varpi_A)
\Big\} 
\notag \\ & 
- e^{- \ImUnit \, \Delta E^{(B)} x_0} \, 
\Big\{ I(\xi_{2+},- \varpi_B,\varpi_A) 
- I(\xi_{2-},- \varpi_B, \varpi_A) 
\Big\} 
\Bigg] ,
\label{eqn:CalIEITOKeyInt} 
\end{align} 
where $I(z_0, \alpha, \beta)$ is an integral defined by 
\begin{equation}
I(z_0, \alpha, \beta) \equiv \int^{\infty}_0 d t \, \frac{1}{t - z_0} \, t^{\ImUnit \beta} e^{\ImUnit \alpha t} ,
\label{eqn:KeyIntforCalIE} 
\end{equation}
$z_0$ is complex, $\alpha$ and $\beta$ are real, and 
\begin{equation} 
\varpi_I \equiv \frac{\Delta E^{(I)}}{\kappa} , \quad 
\xi_{1\pm} = \kappa x_0 \pm \ImUnit \sqrt{1 - \kappa^2 x_0^2} 
, \quad 
\xi_{2\pm} = - \kappa x_0 \pm \ImUnit \sqrt{1 - \kappa^2 x_0^2}  . 
\end{equation}
The integral $I(z_0, \alpha, \beta)$ is implemented by rotating the integration path to the positive or the negative imaginary axis depending on the sign of $\alpha$ being positive or negative, as 
\begin{equation}
I(z_0, \alpha, \beta) = e^{\ImUnit \alpha z_0} \, \left[ ( - z_0 )^{\ImUnit \beta} \, \Gamma(\ImUnit \beta + 1) \, \Gamma(- \ImUnit \beta, \ImUnit \alpha z_0) + z_0^{\ImUnit \beta}  \, C \right] , 
\label{eqn:KeyIntSummary} 
\end{equation}
where $\Gamma(s)$ is the Gamma function and $\Gamma(s,z)$ is the upper incomplete Gamma function, $C$ is given as 
\begin{equation}
C \equiv \left\{ \begin{array}{lll} 
\pm 2 \, \pi \, \ImUnit & \quad & \mathrm{Re} \, z_0 > 0 \; \mathrm{and} \; \pm \mathrm{Im} \, z_0 > 0 \\ 
0 & \quad & \mathrm{otherwise} 
\end{array} \right. ,
\label{eqn:IntAComplexEvalSubC} 
\end{equation} 
and the upper sign in Eq. (\ref{eqn:IntAComplexEvalSubC}) refers to $\alpha > 0$ and the lower sign to $\alpha < 0$. 
From Eqs. (\ref{eqn:CalIEITOKeyInt}) and (\ref{eqn:KeyIntSummary}), $\mathcal{I}_E$ is found to be given by 
\begin{align} & 
\mathcal{I}_E 
= \frac{1}{4 \pi} \frac{1}{\sqrt{1 - \kappa^2 x_0^2}} 
\Bigg[ 2 \pi \ImUnit \, e^{- \Upsilon_B} 
+ e^{- \varpi_A \pi} \Gamma(\ImUnit \varpi_A + 1) \left\{ e^{- \Upsilon_B} \Gamma(- \ImUnit \varpi_A, \ImUnit e^{\ImUnit \varphi_0} \varpi_B) 
- e^{\Upsilon_B}  \Gamma(- \ImUnit \varpi_A, \ImUnit e^{- \ImUnit \varphi_0} \varpi_B) \right\} 
\Bigg]  ,  
\label{eqn:CalIEAABIFin} 
\end{align} 
where 
\begin{equation}
\Upsilon_B \equiv \frac{\Delta E^{(B)}}{\kappa} \sqrt{1 - \kappa^2 x_0^2} + \frac{\Delta E^{(A)}}{\kappa}\varphi_0 , 
\label{eqn:UpsilonBDef} 
\end{equation} 
and $\varphi_0 \equiv \arccos \kappa x_0$. 

On the other hand, the excitation probability $c^2 \mathcal{P}_B$ of Bob in the inertial motion identically vanishes when we exactly set as $\chi_B(\tau_B) = 1$, as expected, since an inertial detector should not register any quanta 
in the Minkowski vacuum. However, this is not sufficient for the following purpose. Rather, we need to take into account explicitly the effects of a long but finite interaction time and adiabatic switching, and derive non-vanishing contribution to $\mathcal{P}_B$. To do so, we assume the form of the switching function $\chi_B(\tau_B)$ as  
\begin{equation}
\chi_B(\tau_B) 
\equiv \tanh \left[ \sigma \left( \tau_B +T \right) \right] 
- \tanh \left[ \sigma \left( \tau_B - T \right) \right] ,   
\label{eqn:SwitchingFuncTanh} 
\end{equation} 
where the positive constants $1 / \sigma$ and $2 T$ denote the timescales of the switching and the interaction time, respectively. When the adiabatic limit of the switching $\Delta E^{(B)} / \sigma \gg 1$ is considered, $\mathcal{I}_B$ defined by Eq. (\ref{eqn:CalIIDef}) and related with $\mathcal{P}_B$ by Eq. (\ref{eqn:IntFactorsPE}) is found \cite{KogaKM18} to be approximated as 
\begin{equation}  
\mathcal{I}_B \simeq e^{- \, \frac{\Delta E^{(B)}}{\sigma} \, \pi} \left[ \frac{2}{\pi^2} 
- 2 \frac{\pi^2 - 4 \, \sigma^2 \, T^2}{( \pi^2 + 4 \, \sigma^2 \, T^2 )^2} \, 
\cos \left( 2 \, \Delta E^{(B)} \, T \right) 
+ \frac{8 \, \pi \, \sigma \, T}{( \pi^2 + 4 \, \sigma^2 \, T^2 )^2} \, 
\sin \left( 2 \, \Delta E^{(B)} \, T \right) \right] ,  
\label{eqn:CalIAdiabaitc} 
\end{equation}  
which vanishes in the adiabatic limit $\Delta E^{(B)} / \sigma \rightarrow \infty$, as it should. 
If, in addition, when the infinitely long interaction time $T \rightarrow \infty$ is considered, we obtain the non-vanishing contribution to $\mathcal{I}_B$ as  
\begin{equation}
\mathcal{I}_B \simeq \frac{2}{\pi^2} \: e^{- \, \frac{\Delta E^{(B)}}{\sigma} \, \pi} . 
\label{eqn:CalIIAdInf} 
\end{equation}

\section{Entanglement and teleportation fidelity} 

Now we analyze whether the pair of Unruh-DeWitt detectors are entangled at the infinite future $t  \rightarrow \infty$. 
For a two-qubit state, as in the present case, the necessary and sufficient condition for the state to be entangled has been shown to be provided by positive partial transpose (PPT) criterion \cite{Peres96,HorodeckiHH96}, which states that a two-qubit state is entangled if and only if the partial transpose of the density matrix of the two-qubit state, i.e., the density matrix transposed with respect to either Alice or Bob only, has a negative eigenvalue. 
When applied to the density matrix Eq. (\ref{eqn:DensityMatrixAB}), the PPT criterion shows \cite{KogaKM18} that 
$\rho_{AB}$ is entangled, if either 
\begin{align}
& \mathcal{P}_A \, \mathcal{P}_B < \left| \mathcal{E} \right|^2 , 
\label{eqn:EntanglementCond1} 
\\ & \qquad \mathrm{or} 
\notag \\ 
& \mathcal{X} < \left| \mathcal{P}_{AB} \right|^2 ,   
\label{eqn:EntanglementCond2} 
\end{align} 
holds, which are exclusive with each other, 
where $\mathcal{X}$ is defined as 
\begin{equation} 
c^4 \, \mathcal{X} \equiv \bra{E_1^{(A)}} \bra{E_1^{(B)}} \rho_{AB} \cket{E_1^{(A)}} \cket{E_1^{(B)}} ,  
\label{eqn:CalXDef} 
\end{equation}  
and hence it is a contribution in order of $c^4$. Furthermore, it has been shown \cite{KogaKM18} that the standard quantum teleportation is not possible under the entanglement condition Eq. (\ref{eqn:EntanglementCond2}) to order of $c^2$. Thus, we will focus on the condition (\ref{eqn:EntanglementCond1}) in what follows. 
Eq. (\ref{eqn:EntanglementCond1}) provides a clear picture of entanglement extraction from the vacuum. The Unruh-DeWitt detectors are entangled when the cross-correlation $\mathcal{E}$ between the detectors larger than the excitation probability $\mathcal{P}_I$ is generated. 

From Eq. (\ref{eqn:IntFactorsPE}), the entanglement condition Eq. (\ref{eqn:EntanglementCond1}) is described in terms of $\mathcal{I}_I$ and $\mathcal{I}_E$, as 
$\left| \mathcal{I}_E \right|^2 - \mathcal{I}_A \, \mathcal{I}_B > 0$, which is rewritten, when Eqs. (\ref{eqn:CalPAAABI}) and (\ref{eqn:CalIIAdInf}) are substituted, as 
\begin{equation} 
\left| \mathcal{I}_E \right|^2 - \frac{2 \Delta E^{(A)}}{\pi^3} \, \frac{e^{- \, \frac{\Delta E^{(B)}}{\sigma} \, \pi}}{e^{2 \pi \frac{\Delta E^{(A)}}{\kappa}} - 1} T > 0 , 
\label{eqn:EntangleCndAABI}
\end{equation} 
where $T \rightarrow \infty$ and $\sigma \rightarrow 0$ are understood for an infinitely long interaction time and adiabatic switching. 
Since $0 < x_0 < \dfrac{1}{\kappa}$, and $\varpi_A$ and $\varpi_B$ are positive definite, we find that $\mathcal{I}_E$ in Eq. (\ref{eqn:CalIEAABIFin}) is analytic from the behavior of the Gamma function $\Gamma(s)$ and the incomplete Gamma function $\Gamma(s,z)$. 
In addition, using the asymptotic form $\Gamma(s,z) \sim z^{s-1} \, e^{-z}$ as $\left| z \right| \rightarrow \infty$, as well as the formula of the absolute value of the Gamma function $\left| \Gamma(\ImUnit x) \right| = \sqrt{\pi / [ x \sinh (\pi x ) ]}$ for real $x$, Eq. (\ref{eqn:CalIEAABIFin}) 
with fixed $x_0$ gives 
\begin{equation}
\left| \mathcal{I}_E \right| \rightarrow \frac{1}{2 \sqrt{\pi}} \frac{1}{\varpi_B} e^{- \varpi_A \frac{\pi}{2} } \sqrt{\frac{\varpi_A}{\sinh ( \varpi_A \pi )}} 
\quad \mathrm{as} \quad \varpi_B \rightarrow \infty 
\label{eqn:CalIEAbsVarPiInf} 
\end{equation} 
and hence $\mathcal{I}_E$ is found to be bounded. We also note that Eq. (\ref{eqn:CalIEAbsVarPiInf}) shows $\left| \mathcal{I}_E \right| \rightarrow 0$ in the limit $\kappa \rightarrow 0$, i.e., when both $\varpi_B$ and $\varpi_A $ go to infinity. Thus, two inertial ($\kappa \rightarrow 0$) detectors , which are infinitely far apart, are not entangled, as expected. 

Since $\left| \mathcal{I}_E \right|$ is bounded, we see from Eq. (\ref{eqn:EntangleCndAABI}) that for a fixed $\sigma$, the two Unruh-DeWitt detectors are not entangled in the limit of the infinite interaction time $T \rightarrow \infty$. However, for a chosen $T$ sufficiently large, one can switch the detector more adiabatically by making $\sigma$ smaller, which exponentially suppresses the second term in Eq. (\ref{eqn:EntangleCndAABI}), in contrast to the linear dependence on $T$. Therefore, even if the detectors interact with the scalar field for a sufficiently long time, they can be entangled by performing the switching adiabatically enough. However, this entanglement extraction is understood with the proviso that the entanglement is ``fragile''  in the sense that it is easily broken, depending on the adiabaticity of the switching.   

If entangled, however, it is meaningful here to see the behavior of the entanglement near the Rindler horizon in the present case also, because ``shadow'' or ``sudden death'' of entanglement near the horizon has been pointed out in the literature \cite{HerdersonHMSZ18,TjoaMann20,GallockYoshimuraTM21-}. This is of particular interest, since Hawking radiation consists of quanta that graze the horizon. In the present case, Alice's closest approach to the Rindler horizon of Bob occurs at $\bar{x}_A = \kappa^{-1}$, and hence Alice's worldline gets closer to the Rindler horizon as $\kappa$ increases. Letting $\kappa$ be large enough, and hence $\varpi_I \ll 1$, while fixing $\kappa \, x_0 < 1$ in order for Alice and Bob not to collide, we find, by using the behavior $\Gamma(s,z) \sim \Gamma(s) - z^s / s$ of the incomplete Gamma function as $z \rightarrow 0$, 
that $\mathcal{I}_E$ is approximated as 
\begin{equation}
\mathcal{I}_E = \frac{1}{2 \pi} 
\Bigg[ \ImUnit \frac{\pi - \varphi_0}{\sqrt{1 - \kappa^2 x_0^2}} 
- \ImUnit \frac{\Delta E^{(B)}}{\Delta E^{(A)}} 
- \Gamma(\ImUnit \varpi_A) \varpi_B^{1 - \ImUnit \varpi_A} \Bigg] + O(\varpi_I^2) . 
\label{eqn:CalIEInfAccel} 
\end{equation} 
Since the absolute value of  the last term in the bracket in Eq. (\ref{eqn:CalIEInfAccel}) is found, by using again the formula for $\left| \Gamma(\ImUnit x) \right|$ for real $x$, to be given as $\varpi_B / \varpi_A = \Delta E^{(B)} / \Delta E^{(A)}$ in the limit $\varpi_I \rightarrow 0$, $\left| \mathcal{I}_E \right|$ is bounded for large $\kappa$ without depending on $\kappa$, when we fix $\kappa \, x_0$. On the other hand, the second term in Eq. (\ref{eqn:EntangleCndAABI}) is linear in $\kappa$, when $\kappa$ is large. Therefore, we find there should exist a critical value of $\kappa$, above which the pair of Unruh-DeWitt detectors are not entangled. Note that the proper distance between Alice and Bob is given as $\kappa^{-1} - x_0$ at the closest approach, and that it decreases as $\kappa$ increases when $\kappa x_0$ is fixed. Therefore, in contrast to a possible  expectation based on the behavior of the Feynman propagator that the two detectors will be more entangled as they approach closer, we find that the existence of the horizon has the stronger effect to inhibit them from being entangled. 

However, there still remains the possibility that the pair of Unruh-DeWitt detectors are entangled, by performing the switching adiabatically enough and letting Alice stay away from the horizon. We then turn to the issue whether one can perform quantum teleportation efficiently based on the entanglement between the two Unruh-DeWitt detectors, $A$ and $B$, extracted from the Minkowski vacuum. Here we focus on the standard scheme of quantum teleportation, with which an unknown quantum state $\cket{\phi}$ of an another qubit $C$ is teleported from Alice to Bob. To do this, Alice first performs the Bell measurement on the subsystem composed of $C$ and $A$, which makes the combined state of $C$ and $A$ one of the four Bell states. Then, Alice sends the outcome $k$ of the Bell measurement through classical communication to Bob, and Bob applies the unitary operation $U_k$ on his qubit $B$ depending on the outcome $k$, which yields a state $\rho_{k}$ at his hand. If the two qubits $A$ and $B$ are maximally entangled, Bob can completely recover the state $\cket{\phi}$ as $\rho_k = \cket{\phi}\bra{\phi}$, by applying a suitable unitary operation $U_k$ in accord with the outcome $k$ sent from Alice. 
This is expressed as the fidelity $\tr \left( \rho_k \, \cket{\phi} \bra{\phi} \right) = \bra{\phi} \rho_k \cket{\phi}$ being unity when $A$ and $B$ are maximally entangled. The probability $p_k$ to obtain any outcome $k$ is found to be $1/4$ equally in the maximally entangled case, and then the fidelity $\sum_k p_k \tr \left( \rho_k \, \cket{\phi} \bra{\phi} \right)$ averaged over all outcome $k$ is unity, as well. 

This would be true in the present case, if two Unruh-DeWitt detectors were maximally entangled. To see this, we write the state of the qubit $C$ as $\cket{\phi} = \alpha \cket{0} + \beta \cket{1}$, where $\left| \alpha  \right|^2 + \left| \beta \right|^2 = 1$ and $\left\{ \cket{0} , \cket{1} \right\}$ is a basis in the Hilbert space of the qubit $C$, and we suppose that the Hilbert spaces of $C$ and $B$ are unitarily related. For example, if the qubit $C$ also is a two-level Unruh-DeWitt detector, the ground state $\cket{1}$ of the qubit $C$ is naturally identified with the ground state 
$\cket{E^{(B)}_0}$ of $B$, and the excited state $\cket{0}$ with $\cket{E^{(B)}_1}$ \footnote{This identification is motivated by the standard notation in quantum information theory, $\sigma_z \cket{0} = \cket{0}$ and $\sigma_z \cket{1} = - \cket{1}$.}. Although this is not necessarily the case, much like the Wigner rotation in the case of moving spins \cite{TerashimaUeda03}, as long as the unitary transformation between the two Hilbert spaces is known, say, as $\cket{\phi}_B = U_0 \cket{\phi}_C$, one can absorb it into Bob's unitary operation as $U_k \rightarrow U_ 0 \, U_k$, without changing the fidelity. Thus, the fidelity is {\it not} reduced due to the Unruh effect, in contrast to the case considered in Ref. \cite{AlsingMilburn03}, if Alice and Bob shared the maximally entangled state. Although the energy eigenstates of $A$ and those of $B$ are defined by the different proper times, one with respect to the inertial observer and the other with respect to the accelerated observer, the standard teleportation is performed with the perfect fidelity, if maximally entangled.  

However, in the present case, the entangled state between Alice and Bob is actually mixed. Then, the fidelity will depend on the state $\cket{\phi}$ and will be reduced by the noise. When Alice and Bob share a mixed state,   
the fidelity $f$ averaged over the state $\cket{\phi}$ is then considered \cite{MassarPopescu95,Gisin96,HorodeckiHH96b} as 
\begin{equation}
f = \int d M(\phi) \sum_k p_k \tr \left( \rho_{k} \cket{\phi} \bra{\phi} \right) , 
\end{equation}
where $d M(\phi)$ is the unitary invariant normalized measure of the state $\cket{\phi}$. 
The maximal value $f_{\mathrm{max}}$ of the averaged fidelity $f$ optimized over the unitary transformation $U_k$ by Bob has been derived in the general context in Ref. \cite{HorodeckiHH96b}, and it has been applied in Ref. \cite{KogaKM18} to the density matrix $\rho_{AB}$ in Eq. (\ref{eqn:DensityMatrixAB}), which yields 
\begin{equation}
f_{\mathrm{max}}(\rho_{AB}) = \frac{2}{3} \left( 1 + c^2 \left[ \left| \mathcal{E} \right| - \frac{\mathcal{P}_A + \mathcal{P}_B}{2} \right] \right) + O(c^4) . 
\label{eqn:TelFidelStand} 
\end{equation} 
under  the entanglement condition Eq. (\ref{eqn:EntanglementCond1}). 
Although $f_{\mathrm{max}}(\rho_{AB})$ in Eq. (\ref{eqn:TelFidelStand}) is positive, it does not necessarily mean that the quantum teleportation is useful in transmission of  the information. In order to see the supremacy of the quantum teleportation, we need to compare it with the fidelity of channels without entanglement. 
The maximal value of the fidelity achievable without entanglement has been found to be $2 / 3$, which is derived by allowing for any measurement on a qubit  \cite{Popescu94,MassarPopescu95,HorodeckiHH99}. Thus, the quantum teleportation is useful, i.e., more efficient than channels without entanglement, when the fidelity $f_{\mathrm{max}}(\rho_{AB})$ exceeds $2 / 3$. From Eq. (\ref{eqn:TelFidelStand}), we immediately see that it occurs if $\left| \mathcal{E} \right| - \left( \mathcal{P}_A + \mathcal{P}_B \right) / 2 > 0$. However, as we have seen above, $\left| \mathcal{I}_E \right|$ and hence $\left| \mathcal{E} \right|$ is bounded, whereas $\mathcal{I}_A$ and hence $\mathcal{P}_A$ linearly diverges in $T$ formally, while $\mathcal{P}_B$ remains very small for adiabatic switching. Therefore, we see that one cannot perform quantum teleportation superiorly compared to channels without entanglement, when we consider the case where the Unruh-DeWitt detectors interact with the quantum scalar field for an infinitely long time with sufficiently adiabatic switching. 

\section{Conclusion and discussion} 

To summarize, we first computed the entanglement extracted from the Minkowski vacuum to the pair of two-level Unruh-DeWitt detectors, one of which is inertial and the other is accelerated. We assumed that they are initially prepared in the ground states at the infinite past and interact with a neutral massless scalar field for an infinitely long time with the switching of the detectors performed sufficiently adiabatically at the infinite past and future. We found, within the perturbation theory, that the extracted entanglement at the infinite future is very fragile, in the sense that it depends on how adiabatically the detectors are switched on and off. 
We saw also that, as in the cases of Gaussian switching in a black hole spacetime \cite{HerdersonHMSZ18,TjoaMann20,GallockYoshimuraTM21-}, even if the detectors are entangled away from the Rindler horizon, they are suddenly disentangled as the worldline of the accelerated detector approaches the Rindler horizon. 

However, there remains the possibility that the two Unruh-DeWitt detectors are entangled, if the switching is performed adiabatically enough and the accelerated detector is away from the Rindler horizon. We then considered the standard scheme of quantum teleportation utilizing the extracted entanglement in that case. We found that the maximal value of the averaged fidelity of the quantum teleportation does not exceed the value achievable without entanglement. Therefore, we concluded that the standard quantum teleportation beyond the horizon, using the entanglement extracted from the Minkowski vacuum only, is not useful.  

If we consider more general schemes of quantum teleportation, the fidelity may exceed the value achievable without entanglement. It is then interesting to ask what happens  if it is indeed the case, not only for the Rindler horizon but also for a black hole horizon. 
If the fidelity of the quantum teleportation is larger than that achievable without entanglement but allowing for any measurement on a qubit emitted as Hawking radiation, can we retrieve the information {\it teleported}, rather than gravitationally fallen,  into a black hole before it has evaporated? 
It will then be interesting to clarify circumstances under which general quantum teleportation beyond a black hole horizon is implemented superiorly.  
We may need to analyze not only the average fidelity of quantum teleportation, but also the deviation of the fidelity \cite{BangRK18} in that case. 

To study this issue further, we also need to clarify the behavior of the entanglement between the inside and the outside of a black hole. 
The results in the literature so far \cite{HerdersonHMSZ18,TjoaMann20,GallockYoshimuraTM21-} and in this paper suggest that the static observer near a black hole horizon is hard to be entangled with a freely falling observer. 
On the other hand, in a moving mirror model, Hawking particles (Rindler particle in the Minkowski spacetime) has been shown to be entangled with Milne particles \cite{Wald19}. It would be interesting to find out whether a pair of Unruh-DeWitt detectors can probe this entanglement. Unfortunately, however, it does not seem probable from our previous analysis based on the energy conservation in Ref. \cite{KogaMK19}. Since both Hawking particles and Milne particles are in the thermal state in the Minkowski vacuum, we expect the behavior of the excitation probability $\mathcal{P}_I$ will grow linearly in time also in this case. Then, in order for the detectors to be entangled, i.e., for the cross-correlation $\mathcal{E}$ to dominate $\mathcal{P}_I$, the worldlines of the detectors would need to be tangent to the same timelike Killing vector field for the energy conservation to hold. However, Milne particles are defined with respect to a conformal Killing vector, not a Killing vector. Hence, a quantum emitted by virtual de-excitation of one detector is not resonantly absorbed into the other detector to virtually excite it, which will suppress $\mathcal{E}$, and thus entanglement will not be extracted. 

In view of the formal divergence linear in time of the excitation probability of Alice, non-perturbative analysis will be desirable.  
Recently, based on the algebraic quantum field theory, it is found that entanglement cannot be harvested, if the detectors are coupled to a quantum field only within acausally separated compact regions in a spacetime \cite{Ruep21}, where the switching functions thus should have compact supports in the proper time. While the initial state of the detectors are forced to be mixed due to the restriction to compact regions in the formalism, this will be helpful in investigations of entanglement extraction in an evaporating black hole spacetime, since Bob will hit the singularity within his finite proper time. 
It will be interesting if we can extend this line of research in the algebraic quantum field theory to include the case of  infinitely long interaction, as well. 

Although we focused in this paper on the ground states as the initial state of the Unruh-DeWitt detectors, it will be meaningful to extend the investigation to the cases of the general initial state of the detectors, including entangled states. If the entanglement between a freely falling detector and a static detector in a black hole spacetime will be preserved or increase, it is interesting to consider quantum teleportation and its possible competition with retrieval processes of the information from a black hole. Quantum teleportation involves classical communication as a part of its task, and thus it will feed a black hole with some amount of energy if the classical signal is sent through a high frequency channel, while low frequency signals will be scattered off by the curvature of a black hole spacetime, which reduces the efficiency of quantum teleportation. Then, it will be interesting to consider the relation between energy and information obtained by a black hole, and thus extend the framework of black hole thermodynamics to include quantum teleportation. 
On the other hand, if the initial entanglement will be lost or degraded, it will be important to ask where it is transferred. 

\acknowledgments 
We would like to thank N. Iizuka, Gen Kimura, and Kei Matsuura for useful discussion. 
This work was supported in part by JSPS KAKENHI Grant Number 17K05451 and 20K03975. 

\baselineskip .2in

\end{document}